\documentstyle[prb,aps]{revtex}
\begin{document}
\draft

 \title{Spin relaxation  in semiconductor quantum dots}

 \author{Alexander V. Khaetskii \cite{address} and Yuli V. Nazarov}
\address{ Faculty of Applied Sciences and DIMES, Delft University of
Technology,\\
Lorentzweg 1, 2628 CJ Delft, The Netherlands}

\maketitle

\begin{abstract}
We have studied the physical processes responsible for the spin -flip in GaAs quantum dots. 
We have calculated the rates for different mechanisms which are related to spin-orbit coupling and cause a spin-flip 
during the inelastic relaxation of an electron in the dot both with and without a magnetic field. 
We have shown that  the zero-dimensional character of the problem when electron wave functions are localized
in all directions  leads to freezing out of the most effective spin-flip
mechanisms related to the absence of the inversion centers in the elementary crystal cell and at
 the heterointerface and, as a result, to  unusually low spin-flip rates.  

\end{abstract}


\par
 Quantum dots (QD) are small conductive regions in a
semiconductor that contain a tunable number of electrons. The shape and size
of  quantum dots can be varied by changing the gate voltage. Besides, the
electronic states can be significantly modified by a magnetic field applied
perpendicular to the plane of the dot. It gives a unique opportunity to study
the properties of the electron quantum states in detail and manipulate with
the electrons in these artificial atoms in a controllable way (see reviews
\cite{1,2}).
\par
As it is known \cite{diVincenzo}, quantum dots are considered  as possible candidates for building a quantum computer. The crucial point of the idea is the necessity to couple dots coherently and  keep coherence on sufficiently long time scales. 
In this respect, there is a great demand in a theoretical estimation of the typical dephasing (especially spin dephasing) time of the electron in the QD.
 This situation can occur, for example, when  the ground state
of the two electron system corresponds to a singlet state $S=0$ and an excited state when one of the electrons occupies the higher orbital level corresponds to
the total spin of the system $S=1$. 
Then an inelastic relaxation to the  ground state of the dot
has to be accompanied by a spin flip. As a result, the spin degree of freedom must be connected through some mechanisms to the external  environment.
Moreover, the electron-electron interaction which can be quite important in determining the energy levels of the system, is not important at all for the spin-flip process itself. Therefore, we can consider the spin-flip processes within the one-electron approach.

\par

We have studied the physical processes responsible for the spin-flip in GaAs quantum dots. The unit cell in this material has no inversion center which gives rise to an effective spin-orbit coupling in the electron spectrum. It is known \cite{Pikus,Dyak}, that such
 spin-orbit coupling provides the main source of spin-flip both in 3D and 2D cases. Besides, in the polar-type crystal there is a strong coupling of electrons to the bosonic environment 
through the piezo-electric interaction with acoustic phonons. Such coupling can be important for the inelastic electron relaxation in the GaAs crystal both with and without a spin-flip.
\par
We have calculated the rates for different spin-orbit related mechanisms which cause a spin-flip 
during the inelastic relaxation of the electron in the dot both with and without a magnetic field. The rates are obtained
as a function of energy transfer. It is generally observed that the corresponding spin-flip rates are by several orders of magnitude lower than 
for any (elastic and inelastic) spin-flip processes for free 2D electrons, i.e. in the case when there is no confinement in the plane.
The reason for that is as follows.  
First of all, some spin-flip mechanisms that are effective in the 3D case
are frozen out in 2D case.
It is known that in the 3D case                 
the spin-flip mechanism related to the spin-orbit admixture of the p-band functions                to the conduction band functions (the so-called Yafet-Elliot                        mechanism \cite{Pikus}) is possible even in  materials with a bulk inversion 
center.
However, in the 2D                   
case the   corresponding                    
scattering amplitude may involve only the following term                        
responsible for the spin-flip: $\hat{\mbox{\boldmath $\sigma$}}\left                  [{\bf p}\cdot                                                                  
{\bf p}'\right]\propto \hat\sigma_z$, where ${\bf p},                                {\bf p}'$ are the two-dimensional electron momenta                              
prior to and after scattering (the z-axis coincides with the normal to the 2D plane). Since $\hat\sigma_z$ contains                        
only diagonal components, this mechanism is suppressed in the 2D case.
Therefore,  in the GaAs-like crystal the most effective mechanism of
the spin-flip in the 2D case                     is related to broken                     inversion symmetry either in the elementary crystal                 
cell or at the heterointerface \cite{Dyak,Rashba}.
This brings about terms in the Hamiltonian \cite{Dyak} 
which are   linear in the two-dimensional electron momentum
and proportional to the 
first power of the small parameter $\Delta/E_g \ll 1$, where
$\Delta$ is the spin-orbit       
splitting of the valence band of the bulk GaAs crystal and $E_g$ is the energy gap.
 This leads to
the quadratic dependence of the spin-flip rate on this parameter in the 2D case
\cite{Dyak}.
\par
 In this paper we show that in the case of the quantum dot the zero-dimensional character of the problem
leads to the additional suppression 
of the spin-flip rate. Namely, the localized character of the electron wave
functions in the dot leads to a fairly  interesting fact that  linear in
 spin-orbit parameter $\Delta$ terms in the matrix elements are absent in true
two-dimensional case.
As a result, the spin-flip rate is proportional to  the power of the parameter $\Delta$ which is
higher than the second one. The contribution to the spin-flip rate which is
quadratic with respect to $\Delta$ exists
only if we take into account the admixture of the electron wave functions of the higher
levels of the size quantization in the z-direction, i.e. the weak deviation
from the true 2D motion. This contribution is also small because
we consider the case of a strong transverse quantization when the lateral electron kinetic energy is much smaller than the distance between the quantized levels in the z-direction.

\par

 We start with the following effective- mass Hamiltonian which is derived
from the Kane model (see Ref. \onlinecite{Pikus})  
         and describes  the
 electron in the conduction band in the presence of  a magnetic field ${\bf B}$ parallel to the z-axis (normal to the 2D plane) and an arbitrary confining potential $U({\bf r})$, both vertical and lateral:
   \begin{eqnarray}                                                 
\hat {\cal H}=\hat {\cal H}_0+\hat {\cal H}_1 + \hat {\cal H}_{ph}
+\hat {\cal H}_{e-ph};                             
    \hat {\cal H}_0=\frac{\hat{\bf p}^2}{2m}+ U({\bf r});  \nonumber \\
                                   \hat {\cal H}_1= 
\frac{\hbar\Delta}{3 m E_g^2}
\hat {\mbox{\boldmath $\sigma$}}  [{\mbox{\boldmath $\nabla$}}U \cdot \hat {\bf p}]
+ \frac{2\Delta}{3\sqrt{2m E_g} m_{cv}E_g} 
{\hat{\mbox{\boldmath $\sigma$}}}   
\cdot {\hat {\mbox{\boldmath $\kappa$}}}
 + \frac{1}{2} V_0  {\hat{\mbox{\boldmath $\sigma$}}}   
\cdot {\hat {\mbox{\boldmath $\varphi$}}} +\frac{1}{2}g\mu_B \hat\sigma_z B,
 \label{1}                          
   \end{eqnarray}                         
Here $\hat {\bf p}=-i\hbar {\mbox{\boldmath $ \nabla$}} +(e/c) {\bf A} $ 
  is the 3D  electron  momentum operator $(e>0)$, $m$  the effective mass, 
 $\hat  {\mbox{\boldmath $\sigma$}} $ 
the Pauli matrices,  $ m_{cv} $
the parameter of the Kane model with the dimensionality of mass which determines the "interaction" 
of the conduction and the valence bands with  other bands, $ \hat\kappa_x = \hat p_x
(\hat p_y^2 -\hat p_z^2)$ and other components  are obtained by cyclic permutation of indices, $x,y,z$ are directed along the main  crystallographic axes. This Hamiltonian gives very accurate description of the conduction band states in the case $E\ll E_g$, where $E$ is the electron energy counted from  the conduction band edge, and all  parameters of this Hamiltonian are very well known for GaAs \cite{Pikus}. 

 The last (Zeeman) term describes the direct interaction of the electron spin with the 
magnetic field, where $g$ is electron g-factor and   $\mu_B$ is the Bohr magneton.
The first and second terms in $\hat {\cal H}_1$ describe   
           the spin-orbit                                                             interaction.  The first term is due to the relativistic interaction with             the electric field caused by the confinement. An impurity potential should be also added to $U$.
           The second term in $\hat {\cal H}_1$ is related to the                          absence of an inversion                                                       
centre in the elementary cell of the GaAs crystal. In the 3D case               
this leads to spin splitting of the conduction band proportional to             
the cube of the electron momentum. In the true 2D case                     
averaging of the corresponding term along the motion of the electron             in the direction perpendicular to the 2D plane gives rise to a spin               splitting proportional to the first power of the 2D electron momentum              \cite{Dyak},                                                              
 this splitting being strongly dependent on the orientation                     
of the vector of the normal to the 2D plane  with       
respect to the main crystal axes.                                   
 Below we shall consider only the case when        the
    normal is parallel to the [100] direction which is the most frequently used orientation.
\par
We have  also included in the above Hamiltonian the spin-orbit splitting of the electron spectrum due to the strain field
produced by acoustic phonons. There $\hat \varphi_x = (1/2)\{u_{xy},\hat p_y \}_+ - (1/2) \{u_{xz},\hat p_z \}_+ $, where  $\{,\}_+$ means anticommutator,
 $\varphi_y, \varphi_z$ are obtained by cyclic permutations, $u_{ij}$
is the lattice strain tensor \cite{Levinson} and $V_0$ is the characteristic velocity whose value is well known for GaAs \cite{Titkov}, 
 $V_0 = 8\times 10^7 cm/s$. 
\par
We have considered  all possible mechanisms of the spin-flip which are related to the spin-orbit coupling. These are: 
\par
1). The spin-flip event is caused by the spin-orbit relativistic coupling to the electric field (polarization) produced by the emitted phonon. 
This process is described by the  term proportional to $V_0$ in Eq.(\ref{1}).

\par
2). The spin-flip  event is caused by  the spin-orbit admixture of different spin states. 
This means that in the presence of the spin-orbit term in the Hamiltonian the electron "spin-up" state contains actually  a small admixture of the  "spin-down" state.  The electron transition with
phonon emission between two states with "opposite" spins leads to the
spin-flip event. This mechanism is related to the first and second terms in the Hamiltonian  $\hat {\cal H}_1$
and the phonon provides  only  energy
conservation here.
\par
3). In  GaAs  the electron $g$-factor ($g=-0.4$) differs  strongly
from the free electron value $g_0=2$ which is due to the  spin-orbit interaction
which mixes  valence band  and conduction band states \cite{Kittel}.
 This leads to a new mechanism of spin-flip in GaAs in the presence of an external magnetic field. Namely, the strain produced by the emitted phonon   is coupled directly to the 
electron spin through the variation of the effective $g$-factor:
$\simeq g\mu_B u_{iz}\hat\sigma_i B_z$.
\par
It is not clear a-priori which mechanism provides the main source of the spin-flip. We found actually that the second mechanism related to the admixture of the different spin states is the most effective one. Below we consider 
these mechanisms one by one.

\par
1). First mechanism. We investigate  only the relaxation of the spin component perpendicular to the 2D plane ($S_z$ component). Then, using the standard expression for the strain tensor due to the acoustic
deformation \cite{Levinson}, we get for the matrix element describing the electron spin-flip transition 
between spatial states $\Psi_1$ and $\Psi_2$
with emission of a phonon:
 
\begin{equation}
M_{\uparrow,\downarrow} =  \frac{V_0}{4}
\left (\frac{\hbar}{2\rho \omega_Q} \right )^{1/2}[q_x e_y + q_y e_x] \left <\Psi_1\mid \frac{1}{2}
\{ (\hat p_x +i \hat p_y), \exp(i{\bf Q r})\}_+ \mid \Psi_2 \right >, 
\label{2}
\end{equation}
where $\rho$ is the crystal mass density, ${\bf e}$ the phonon unit polarization vector, ${\bf Q}=({\bf q}, q_z)$ the phonon wave vector. The spin decay rate is given by the Fermi golden rule:

\begin{equation}
\Gamma _{\uparrow,\downarrow} = \frac{\pi \hbar V_0^2}{16 \rho \epsilon}  \int \frac{d^3 Q}{(2\pi)^3} 
 (q_x^2 + q_y^2) \mid  \left <\Psi_1\mid \frac{1}{2}
\{ (\hat p_x +i \hat p_y), \exp(i{\bf Q r})\}_+ \mid \Psi_2 \right > \mid ^2  \delta (\hbar s Q - \epsilon),
\label{3}
\end{equation}
where $\epsilon$ is the energy transfer. We consider transitions between  neighbouring discrete 
energy levels, so that $\epsilon= \hbar s Q \approx \hbar^2/ m \lambda^2$, where $\lambda$ is the dot size in the lateral directions. On the other hand,  
in practically interesting case of transitions between low-lying energy levels,
the energy transfer is of the order of electron energy itself. This means that
perturbation  theory (including only one-phonon processes) is  valid when $\epsilon \gg m s^2$, $s$ being the sound velocity. From this we obtain the condition $q_z \gg q \approx 1/\lambda$, i.e. the phonon is emitted almost perpendicular to the 2D plane. Then, assuming that $q_z z_0 \ll 1$ , where $z_0$ is the width of the 2D layer in the $z$- direction, we can  easily calculate
the integral over $q_z$:

 \begin{equation}
\Gamma _{\uparrow,\downarrow} = \frac{V_0^2}{32 \rho s \epsilon} \int \frac{d^2 q}{(2\pi)^2} 
 q^2  \mid \int d^2 r \Phi_1 \frac{1}{2} \{\hat p_x + i \hat p_y , \exp(i{\bf q}{\bf r})\}_+ \Phi_2 \mid ^2 = -\frac{V_0^2}{128 \rho s \epsilon} \int
 d^2 r \Phi_{12} ^{\star} \mid  \nabla_- \mid ^2 \Phi_{12}
 \label{4}
\end{equation}
where $\Phi_{12} = \Phi_1 \nabla_- \Phi_2^{\star} - \Phi_2^{\star} \nabla_- \Phi_1$,
 $\nabla_{\pm} = \nabla_x \pm i \nabla_y$ and
$\Phi_1, \Phi_2$ are the lateral electron wave functions.
We give here the final expressions in the particular case of elliptic 
quantum dot with  confining frequencies $\omega_x$ and $\omega_y$. For the transition $n_x =1, n_y =0 \Rightarrow n_x =0, n_y =0 $ 
$(\epsilon = \hbar \omega_x)$  we have:
 
 \begin{equation}
\Gamma _{\uparrow,\downarrow}
 = \frac{m^3 V_0^2 \epsilon^2}{128 \pi \rho s \hbar^4}
 \sqrt{\frac{\omega_y}{\omega_x}} (1 + \frac{\omega_y}{\omega_x}).
 \label{5}
\end{equation}
In the case of transition 
 $n_x =0, n_y =1 \Rightarrow n_x =0, n_y =0 $ 
$(\epsilon = \hbar \omega_y)$ in the above formula $\omega_x $ should be replaced by $\omega_y$ and vice versa .
With the use of the $V_0$ value mentioned above, 
 the corresponding relaxation time is found to be very long 
($10 \div 10^{-1}$ s) for a typical energy transfer of the order of
 $1\div 10 $ K.

\par 2). Second mechanism. We start with the general expression which gives the phonon-assisted transition rate   between two states 1 and 2 (which can have either  the same  or "opposite" spins). We consider here the [100] orientation and take into account only the interaction with piezo-phonons which is known to be the most effective in  polar crystals for a low energy transfer:
 \begin{equation}
\Gamma _{12} = \frac{2\pi}{\hbar}2 \frac{\hbar (eh_{14})^2}{2 \rho s_t}  \int \frac{d^3 Q}{(2\pi)^3} 
 \frac{A_t(Q)}{Q} 
 \mid \int d^2 r  \exp(i{\bf q}{\bf r})\hat\Phi_1^{\dagger} ({\bf r})\hat\Phi_2  ({\bf r})\mid ^2
\delta (\hbar s_t Q - \epsilon),
\label{Fermi}
\end{equation}
where $h_{14}$ is the piezomodulus, $eh_{14} = 1.2\times 10^7 eV/cm$ for GaAs.
 As before, the phonon is emitted almost perpendicular to the 2D electron plane. In the case of the interaction with the piezo-phonon this gives rise to an anisotropy effect. Namely, the transition rate strongly depends on the orientation of the normal to the 2D plane with respect to the main crystallographic axes. The rate is strongly suppressed when the normal is parallel to the [100] axis and this  is taken into account  in Eq.(\ref{Fermi}) by the multiplier $A_t(Q)\approx 2 q^2/q_z^2 \ll 1$  which describes the 
anisotropy effect for a transverse phonon (longitudinal phonons give much smaller contribution). There is no such a small factor for the  [111] orientation and consequently 
the  transition rates for the [100] and [111] orientations of the heterostructure differ by more than an order of magnitude.
Proceeding as before, we obtain from Eq.(\ref{Fermi}):
\begin{equation}
\Gamma _{12} = -\frac{2s_t \hbar^2 (eh_{14})^2}{\rho  \epsilon^3} \int
 d^2 r \chi_{12} ^{\star} \mid  \nabla_- \mid ^2 \chi_{12},
 \label{5.0}
\end{equation}
where $\chi_{12}=  \hat\Phi_1^{\dagger} \hat\Phi_2 $.
For  reference we also give the expression for the transition rate {\it without} spin-flip which follows from Eq.(\ref{5.0}). For the transition $n_x =1, n_y =0 \Rightarrow n_x =0, n_y =0 $ in an  elliptic 
quantum dot
$(\epsilon = \hbar \omega_x)$  we have:
\begin{equation}
\Gamma _0 = \frac{(eh_{14})^2 m^2 s_t}{2\pi\rho \hbar^2 \epsilon}
 \sqrt{\frac{\omega_y}{\omega_x}} (3 + \frac{\omega_y}{\omega_x}),
\label{5.1}
\end{equation}
which for the transfer energy of 1K has the value 
$\approx 5\times 10^8 s^{-1}$. Again, in the case of transition 
 $n_x =0, n_y =1 \Rightarrow n_x =0, n_y =0 $ 
$(\epsilon = \hbar \omega_y)$ in the above formula $\omega_x $ should be replaced by $\omega_y$ and vice versa .

\par
In the true 2D case and for the [100] orientation
  we obtain the following spin-orbit Hamiltonian from Eq.(\ref{1}) :
 \begin{equation}
\hat {\cal H}_1=
\beta (-\sigma_x \hat p_x + \sigma_y \hat p_y); \/
\beta=\frac{2}{3} \langle p^2_z \rangle \frac{\Delta}                           
{(2mE_g)^{1/2} m_{cv} E_g}.    
\label{6}
\end{equation}
Note, that for  the other orientation [111]  used in  practice the spin-orbit Hamiltonian has similar structure (i.e. contains the same Pauli matrices), so
our conclusions presented below also valid for this case.
 The constant $\beta$ in Eq.(\ref{6})              
depends on  the mean value of the $\hat p_z^2$ operator in the state described by the  wave function $\chi_0 (z)$ of the lowest quantized level in the $z$- direction and 
for GaAs heterostructures takes values between
 $(1\div 3) \cdot 10^5 cm/s$. In Eq.(\ref{6}) we have dropped the so-called Rashba term \cite{Rashba} which is believed to be much smaller for GaAs heterostructures (and in any case its presence does not influence our conclusions because it has a similar structure).
The presence of  $\hat {\cal H}_1$  term
leads to a nonzero value of the spin-flip transition matrix element. At first sight, the scalar product of the spinors
$\hat \Phi$ corresponding to the initial spin-up and the final spin-down states should be proportional to the first power of $\beta$. However, in contrast to the  extended 2D states, in quantum dots we can actually remove the terms linear in $\beta$ from the Hamiltonian by the following spin-dependent transformation:

 \begin{equation}
\hat \Phi =
[ \hat I + \frac{im\beta}{\hbar} (x\hat\sigma_x  - y\hat\sigma_y)]\hat \Phi'.
\label{7}
\end{equation}
We stress that the boundedness of the electron wave functions is essential for this procedure.
Then we obtain the Hamiltonian which besides the terms with a unit spin matrix contains only the following spin-dependent term: $(m\beta^2 /\hbar) \hat L_z \hat \sigma_z$, where $\hat L_z = -i\hbar \partial /\partial \varphi$ is the angular momentum operator. Therefore, in this approximation the correct spin functions are the eigenfunctions of the $\hat \sigma_z$ operator and there are no spin-flip processes. The above mentioned scalar product of the spinors corresponding to "opposite" spin directions is proportional to the third power of $\beta$. The corresponding spin-flip rate $\simeq
\Gamma_0 (m\beta^2/\hbar \omega)^3 $ is very small, $\Gamma ^{-1} \simeq 10^{-4} s$ for the energy transfer
of the order of 1K. We give here the result  
 for the case of circular dot ($\omega_x =\omega_y = \omega_0$) for the
 phonon- assisted spin-flip transition between first excited  and ground states:
 \begin{equation}
\Gamma _{\uparrow,\downarrow} = \frac{16(eh_{14})^2 m^2 s_t}{3\pi \rho\hbar^2 \hbar\omega_0} \left (\frac{m\beta^2}{\hbar \omega_0}\right)^3 
 \label{result}
\end{equation}

Note that the term $\propto \hat\sigma_x p_x p_y^2 - 
\hat\sigma_y p_y p_x^2 $ in the Hamiltonian  $\hat {\cal H}_1$ (Eq.(\ref{1})) which is proportional to the third power of the in-plane momentum cannot be removed by the above mentioned transformation and gives the contribution to the value of $\chi_{12}$ , Eq.(\ref{5.0}), in the first order of the perturbation theory. This contribution 
can be comparable to that given by Eq.(\ref{result}).
The result for the transition $n_x =1, n_y =0 \Rightarrow n_x =0, n_y =0 $ in an  elliptic quantum dot $(\epsilon = \hbar \omega_x)$ is:
\begin{equation}
\Gamma _{\uparrow,\downarrow} = \frac{3(eh_{14})^2 m^2 s_t}{2\pi\rho \hbar^2 }
 m\alpha^2 \sqrt{\frac{\omega_y}{\omega_x}} \left [\left (1 +\frac{\omega_x}{\omega_y}\right) \left(\frac{\omega_x +\omega_y}{2\omega_x +\omega_y}\right)^2
+ \left(1+ 5 \cdot\frac{\omega_y}{\omega_x}\right)
\frac{\omega_y^4}{(\omega_x^2 - 4 \omega_y^2)^2 } \right ],
\label{result1}
\end{equation}
where $\alpha = \beta /E_z$, $E_z =  \langle p^2_z \rangle /m$. This contribution  is of the order of $ 
\Gamma_0 (m\beta^2\hbar \omega/E_z^2 )$  and becomes comparable to that given by
Eq.(\ref{result}) at $\hbar \omega \simeq \sqrt{m\beta^2 E_z}$ which is of the order of several K.

\par The structure of the transformed Hamiltonian , i.e. the existence of only the $\hat \sigma_z$ matrix, suggests an interesting anisotropy effect with respect to the direction of the external magnetic field which can be observed for the spin-flip processes. Namely, if the direction of the external magnetic field constitutes some angle $\theta$  with the direction of the normal (z-axis),
then the true spin quantization axis will be determined by the magnetic field direction and the above mentioned spin-orbit term will lead to spin-flip
processes with a rate proportional to $\beta^4$  and angular dependence 
$\sin^2 \theta$.
\par  

There are, of course, contributions to the spin-flip rate proportional
to $\beta^2$ which are either related to the virtual transitions to the higher quantized energy levels in the z-direction or to the presence of an impurity potential which leads to the nonseparability of the transverse (z) and longitudinal variables in the  Hamiltonian (\ref{1}). However, it can be easily checked that both effects give  actually the small contributions to the spin-flip rate.
For example, in the first case  the rate  $\Gamma _{\uparrow,\downarrow} \simeq \Gamma_0 (m\beta^2 \epsilon^3/ms^2 E_z^3)$  is very small ($\simeq 1s^{-1}$), here $E_z$ is the distance between the quantized levels in the z-direction. For the second case
 $\Gamma _{\uparrow,\downarrow} \simeq
\Gamma_0 (m\beta^2/\hbar \omega)(U_{imp}/E_z)^2 (z_0/r_c)^2$, where $\hbar \omega$ is the 
typical distance between the levels in a dot, 
 $z_0$ is the thickness of the 2D layer in the z-direction, $r_c$ is the correlation radius of the donor potential and $U_{imp}$ is a magnitude of this potential which we take to be not larger than  $\hbar \omega$. This rate is 
obviously smaller than that given by Eq.(\ref{result1}).
Finally, in the presence of a magnetic field $B$ directed along the z-axis
 there 
is also a contribution to the spin-flip rate proportional to
$\beta^2$. It appears if one takes into account the finite Zeeman splitting in the energy spectrum. We give here the final expression for the case of circular
dot ($\omega_x =\omega_y = \omega_0$) when the solutions of the $\hat{\cal H}_0$
Hamiltonian are well known. These are Darwin-Fock states which are characterized
by two quantum numbers: $n,l$. Then for the transitions $n=0, l=\pm 1, \uparrow 
\Rightarrow n=0, l=0, \downarrow$ 
we obtain with the use of Eqs.(\ref{5.0},\ref{6}):
\begin{equation}
 \Gamma_{\uparrow,\downarrow} = \frac{12(eh_{14})^2 m^2 s_t \omega^3}{\pi\rho \omega_0 \epsilon^3}
 \left (\frac{m\beta^2}{\hbar \omega_0}\right) \left(\frac{g\mu_B B}{\hbar \omega_0}\right)^2,
\label{7.0}
\end{equation}
where the energy transfer $\epsilon (l=\pm 1) = \hbar \omega + \hbar \omega_c l/2 + g\mu_B B, $ $\omega = \sqrt{\omega_0^2 + (\omega_c^2/4)}$ and 
 $\omega_c = eB/mc$ is the cyclotron frequency.
Note that this rate is of the order of $\Gamma_0 (m\beta^2/\hbar \omega_0) (g\mu_B B/\hbar \omega_0)^2$ and for $l=+1$ equals $5\cdot 10^{+5} s^{-1}$ for $\hbar \omega_0 \simeq 1K$ and $B =1T$  (Zeeman energy $\approx 0.3K$). 
\par
Thus, for   energy transfers  of the order of 1K the  spin-flip time related to the admixture mechanism cannot be shorter than $(10^{-5} \div 10^{-6}) s$.
\par
3). Third mechanism. To describe it we need to calculate the following antisymmetric combination \cite{Kittel} in the presence of a deformation

\begin{equation}
\left <\gamma \mid \hat D^A_{xy} (\hat u) \mid \gamma' \right >=
\frac{\hbar^2}{m_0^2}\left [ \left <\gamma \mid \hat p_x \cdot \frac{1}{\epsilon_c - \hat H (\hat u)} \cdot \hat p_y \mid \gamma' \right >        -  \left <\gamma \mid \hat p_y \cdot \frac{1}{\epsilon_c - \hat H (\hat u)} \cdot \hat p_x \mid \gamma' \right >
\right ]
\label{anti}
\end{equation}
between the states $\gamma, \gamma' = \pm 1/2$ of the conduction band.
Energy $\epsilon_c$ corresponds to the bottom of the conduction band, $m_0$ is the bare electron mass and $\hat p_x, \hat p_y$
are the momentum operators in the absence of the  magnetic field. 
In Eq.(\ref{anti}) we sum over  the valence band  states (the heavy and light holes and the split-off band). The Hamiltonian $\hat H(\hat u)$ (6$\times$6 matrix) describes the valence bands in the presence of strain:
 \begin{equation}
\hat  H (\hat u)=\hat  H (0) + \hat  H_1 (\hat u),
\label{a}
\end{equation}
where $\hat  H (0)$ is the diagonal ($6\times 6$) matrix whose elements correspond to the energy position of the top of the valence  ($-E_g$) and split-off bands ($-(E_g + \Delta)$) and  $\hat  H_1 (\hat u)$ is a matrix   linear in the strain tensor elements which describes the interaction of holes with phonons \cite{Pikus}. This interaction is described by three deformation potential constants  \cite{Pikus} in contrast to only  one constant for conduction band electrons. Expanding Eq.(\ref{anti}) with respect to 
 $\hat  H_1 (\hat u)$, we obtain the following additional term in the Hamiltonian which should be added to Eq.(\ref{1}) and is responsible for the spin-flip in the conduction band:
 \begin{equation}
\hat {\cal H}_{so} (\hat u)=
i \hat D^A_{xy}(\hat u) \frac{e B_z}{\hbar c} \simeq \frac{m_0}{m}\frac{\eta}{(1-\eta /3)} \mu_B \frac{d(u_{xz}\hat \sigma_x + u_{yz} \hat \sigma_y) B_z}{E_g},
\label{b}
\end{equation}
where $\eta = \Delta/(E_g + \Delta) \approx 0.2$ for GaAs and $d$ is one of three deformation  constants \cite{Pikus} which is of the order of several eV.
Making use of the Hamiltonian (\ref{b}) and calculating the spin-flip rate along the steps outlined while deriving Eq.(\ref{5}), we obtain:
 \begin{equation}
\Gamma _{\uparrow,\downarrow} \simeq \left(\frac{\eta}{(1-\eta /3)}\right )^2 
\left (\frac{\hbar \omega_c}{E_g}\right )^2 
\frac{m d^2 \epsilon^2}{\rho s^3 \hbar^4}. 
\label{7.1}
\end{equation}
 This equation corresponds again to the spin-flip transitions between the neighbouring orbital levels.
The rate given by Eq.(\ref{7.1}) (for $\hbar \omega_c \simeq \epsilon$) has the same order of magnitude as that given by Eq.(\ref{5}) and is very small.

\par The results described above are concerned with inelastic transitions between  neighbouring quantized energy levels in a dot which corresponds
 to a relatively large energy transfer. The case of
 spin-flip processes between the Zeeman sublevels of the same orbital level
was considered earlier in Ref. \onlinecite{Frenkel} by D.Frenkel. 
This calculation   considers the spin-flip processes with
phonon emission for electrons in spatially localized states in a quantizing
magnetic field (the quantum Hall effect regime) and deals with only one spin-flip mechanism related to the spin-orbit interaction with the electric field produced by the emitted phonon  (the first one in our classification).
The analog of his result for our case can be easily obtained from Eqs.(\ref{2},\ref{3}), assuming that the energy
transfer is much smaller than the typical distance  $\hbar \omega$ between the levels in 
a dot: $\epsilon = g\mu_B B \ll \hbar \omega$ which in turn means that the 
phonon wave length is larger than the dot size. We consider here only the case of circular dot with confining frequency $\omega_0$ and states with $n=0$ and  $l=0,\pm 1$ (the ground state and first two excited states).
 Then instead of Eq.(\ref{5}) we have:
 \begin{equation}
\Gamma _{\uparrow,\downarrow} =
\frac{V_0^2 (g\mu_B B)^5}{240 \pi \rho s^7 \hbar^4} [l+ \frac{\omega_c}{2 \sqrt{\omega_0^2 + (\omega_c^2/4)}} (\mid l\mid + 1)!]^2, 
\label{8}
\end{equation}

This rate is very small ($\simeq 10^{-2}\div  10^{-1} s^{-1}$) for not too large magnetic field
(order of 1T). The spin-flip rate related to the third mechanism is given by 
$(m_0\eta/m)^2 
 (d/E_g)^2 ((\mu_B B)^5/\rho s^5 \hbar^4) $ and has even smaller value.
We found that the most effective spin-flip mechanism is that related
to the admixture of the different spin states.
As it was mentioned earlier, in the presence of the Zeeman splitting we have the contribution to a spin-flip rate proportional to $\beta^2$.
Then, for example, for the transition between the Zeeman sublevels of the ground state of circular dot with emission of the piezo-
phonon we obtain:
 \begin{equation}
\Gamma _{\uparrow,\downarrow}
=\frac{4}{35\pi}
\frac{(g\mu_B B)^5}{(\hbar \omega_0)^4}\frac{(eh_{14})^2 \beta^2}{\rho \hbar^2}
\left (\frac{1}{s_l^5} + \frac{4}{3s_t^5}\right), 
\label{9}
\end{equation}
where $s_l$ and $s_t$ are longitudinal and transverse sound velocities and while calculating we have used the anisotropy factors for   longitudinal and transverse piezo-modes given in Ref. \onlinecite{Price}.
The ratio of the rates given by Eq.(\ref{9}) and Eq.(\ref{8}) is:
$(eh_{14}/mV_0/\hbar)^2 (m\beta^2 ms^2/(\hbar \omega_0)^4) $
which for $\hbar \omega_0 \simeq 1K$  is of the order of $10^{6}$.
So, the spin-admixture mechanism is much more powerful.
\par
In conclusion, we have calculated the rates for the phonon-assisted spin-flip transitions of an electron in a quantum dot for all possible
mechanisms which are related to the  spin-orbit coupling and provide the connection with the spin degree of freedom. 
We have found unusually low values of the spin-flip rates. 
We  show that it is due to
 the localized
 character of the electron wave functions  in a quantum dot which  leads to freezing out of the most effective intrinsic spin-flip
mechanisms related to the absence of the inversion 
symmetry in the GaAs-like crystals.

\par {\it Acknowledgements.}
This work is  part of the research program of the "Stichting voor Fundamenteel
Onderzoek der Materie (FOM)". We acknowledge support by the NEDO project NTDP-98.
We are grateful to  L.P. Kouwenhoven, T.H. Oosterkamp and G.E.W. Bauer 
for useful discussions.

\end{document}